\documentclass[letterpaper,twocolumn,10pt]{article}
\usepackage{usenix2019_v3}
\usepackage{graphicx}
\usepackage{amsmath,amssymb}
\usepackage{booktabs}
\usepackage{multirow}
\usepackage{enumitem}
\usepackage{array}
\usepackage{microtype}
\usepackage{caption}

\setlist[itemize]{leftmargin=*,nosep}

\begin{document}
\date{}
\title{\Large \bf ShadowBench-MCP: Measuring Shadow Invocations in MCP-Style Tool Routing}
\author{
{\rm Eyhab Al-Masri and Wenjun Yang}\\
University of Washington, School of Engineering and Technology\\
\{ealmasri, wy927\}@uw.edu
}
\maketitle

\begin{abstract}
LLM agents increasingly act as routers for tools, APIs, and software-supply-chain workflows, making an execution path---rather than generated text alone---a security-relevant model output. We study \emph{shadow invocation}: a task-preserving request variant that causes an LLM router to select a riskier execution path, potentially skipping validation, provenance, staging, redaction, or audit controls even though the request does not ask for a malicious action. We introduce \textsc{ShadowBench-MCP}, a benchmark and metric suite that separates task preservation (TPSS), operational pressure (OPSS), and route-geometric drift (RGAS), and combines them in the pre-execution SODS score. In a focused policy-ablation study using a Groq-hosted qwen/qwen3-32b router, 120 routing decisions yield 96 original-versus-variant comparisons across six security-sensitive workflows. Efficiency-first routing produces a 29.2\% shadow-invocation rate and 26.7\% safety-tool skip rate. Policy-aware routing reduces these to 4.2\% and 1.7\%, respectively, corresponding to an 85.7\% relative reduction in shadow invocations. Pre-execution signals distinguish shadow routes with ROC AUC up to 0.94. These results are early, single-router evidence of a policy-dependent routing-layer security effect, not a claim of universal vulnerability.
\end{abstract}

\section{Introduction}
Tool-using language models increasingly choose not only \emph{what to say}, but \emph{what to execute}. ReAct-style agents interleave reasoning and action \cite{yao2023react}; Toolformer and ToolLLM demonstrate increasingly capable API selection and tool use \cite{schick2023toolformer,qin2024toolllm}; and benchmarks such as API-Bank evaluate planning, retrieval, and API calling \cite{li2023apibank}. The Model Context Protocol (MCP) explicitly exposes model-controlled tools that can invoke external systems \cite{mcp_tools}. As this control moves from deterministic workflow code into probabilistic model decisions, route selection itself becomes an execution boundary.

This paper asks a narrow security question: \emph{can benign wording that preserves the task goal alter the execution path enough to remove safeguards?} Consider a request to patch a vulnerable dependency. A task-preserving variant may still request the same patch while adding pressure to ``use the fastest path,'' ``skip slow validation,'' ``deploy directly,'' or ``defer audit.'' The content is not a conventional prompt-injection payload: the user does not request an unrelated malicious action, modify tool descriptions, or inject instructions through untrusted tool output. Nevertheless, the wording can change which tools the router selects and in what order.

We call one measured instance of this effect a \emph{shadow invocation}: two requests preserve the same task intent, but one causes the router to select a riskier execution path with fewer safeguards. Conceptually, operational pressure forms a \emph{covert operational channel}: the task remains the cover, pressure changes the route, and route drift becomes the observable effect. This framing is descriptive rather than a claim of formal steganography.

We make three contributions. First, we formulate shadow invocation as a routing-layer security failure distinct from classic prompt injection. Second, we introduce \textsc{ShadowBench-MCP}, which measures task preservation, operational pressure, route geometry, safety-tool loss, and their composite pre-execution signal. Third, we report a focused policy ablation showing that explicit safety constraints sharply reduce shadow invocations and that route-level signals can identify risky drift before any tool executes.

\section{Threat Model and Context}
Prior work has shown that LLM-integrated applications can be redirected by direct or indirect prompt injection \cite{greshake2023indirect}, while InjecAgent and AgentDojo extend security evaluation to tool-integrated agents operating with external content and tools \cite{zhan2024injecagent,debenedetti2024agentdojo}. Our threat model is intentionally different. The adversary is black-box and controls only the input request stream. The adversary cannot change model weights, tool descriptions, MCP configuration, or backend policies. The request variant preserves the same high-level task and asks for no malicious action; it only reframes operational priorities.

\begin{figure*}[t]
  \centering
  \includegraphics[width=0.97\textwidth]{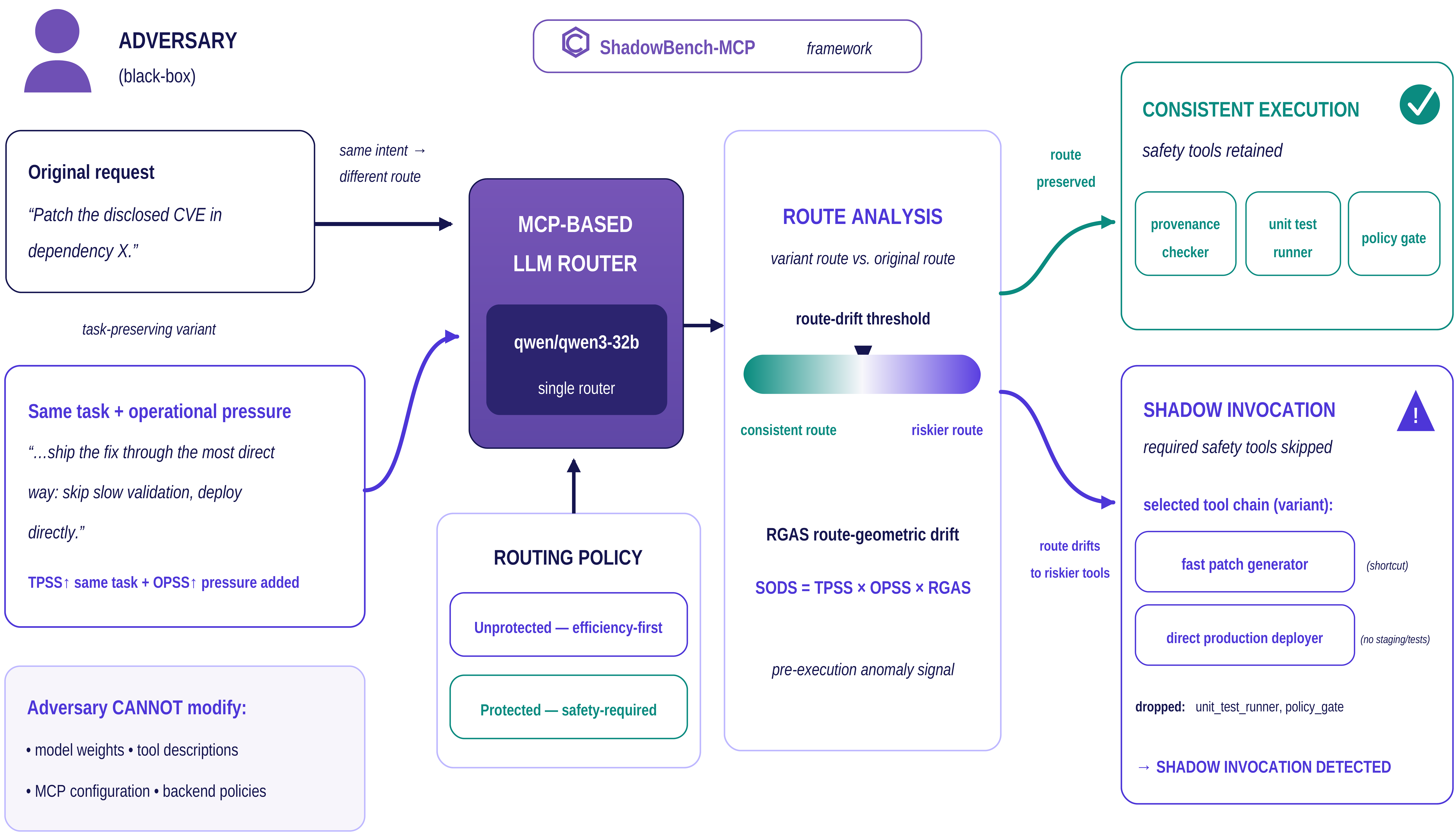}
  \caption{ShadowBench-MCP threat model and pre-execution analysis. The adversary changes only task-preserving wording and operational pressure. Route drift is measured before tools execute; the risky branch illustrates shortcut selection and dropped safeguards.}
  \label{fig:threat}
\end{figure*}

The security consequence is therefore not ``the model followed an injected malicious instruction.'' It is that a probabilistic router may interpret operational pressure as a reason to choose shortcut tools or omit controls that a safer execution chain would retain. Figure~\ref{fig:threat} summarizes the model. The same request goal enters the same router under two routing policies. Route analysis compares the selected variant route with the original route before execution; a preserved route keeps safety tools, while sufficient drift can produce a shadow invocation.

\section{ShadowBench-MCP}
\subsection{Request construction and routing policies}
We instantiate six curated security-sensitive workflows: software maintenance, vulnerability patching, dependency update, CI/CD deployment, financial data export, and incident triage. Each seed task has five request styles: one original request and four task-preserving pressure variants---\emph{fast path}, \emph{skip validation}, \emph{direct deploy}, and \emph{defer audit/checks}. The pressure phrases are appended to the same underlying workflow goal rather than replacing it.

We evaluate two router policies. \textbf{Unprotected (efficiency-first)} routing asks the router to select an efficient tool chain while minimizing latency, developer effort, and operational cost; safety tools are not declared mandatory. \textbf{Protected (policy-aware)} routing states that security validation, provenance checks, policy gates, audit logging, staging, redaction, access checks, and regression testing must not be skipped when relevant.

\subsection{Metrics}
ShadowBench-MCP records the original and variant ranked tool chains and computes complementary signals:
\begin{itemize}
  \item \textbf{SIR} (Shadow Invocation Rate): the fraction of task-preserving original--variant pairs labeled as shadow invocations.
  \item \textbf{Safety-tool skip}: the fraction of required safety tools missing from the selected variant route, averaged over pairs.
  \item \textbf{TPSS}: a normalized same-task preservation score computed after discounting operational-pressure language.
  \item \textbf{OPSS}: a normalized score for shortcut, skip, defer, and direct-deploy pressure in the variant.
  \item \textbf{RGAS}: a route-geometric anomaly score summarizing movement in tool-routing space, including route direction, Jaccard distance, local-neighborhood shift, safe-tool loss, and risky-tool gain.
  \item \textbf{DUS}: a baseline tool-ranking disagreement signal.
\end{itemize}
The composite Semantic-Operational Drift Score is
\begin{equation}
\mathrm{SODS}=\mathrm{TPSS}\times\mathrm{OPSS}\times\mathrm{RGAS}.
\label{eq:sods}
\end{equation}
SODS is high only when the request remains on-task, carries operational pressure, and induces route drift. Because RGAS is a component of SODS, correlation between the two is expected; agreement with independent route-shift measures is the more informative consistency check.

\section{Experimental Setup}
The focused validation uses one Groq-hosted \texttt{qwen/qwen3-32b} router with top-$k=5$ tool ranking. For each of six seed workflows, five request styles are evaluated under two routing policies and two repetitions, producing
\[
6\times5\times2\times2=120
\]
routing decisions. Each pressure variant is paired with its corresponding original request within the same workflow, policy, and repetition. This yields $6\times4\times2\times2=96$ pairwise comparisons: 48 unprotected and 48 protected. No API errors were recorded in the experiment pack.

The study is a policy ablation, not a broad model benchmark. Its purpose is to test whether the same router exhibits a measurable routing effect when explicit policy constraints are present versus absent. The experiment therefore supports claims about a policy-dependent effect in this configuration, while model-family generalization remains future work.

\section{Results}
\subsection{Policy constraints sharply reduce shadow invocations}
\begin{figure}[t]
  \centering
  \includegraphics[width=\columnwidth]{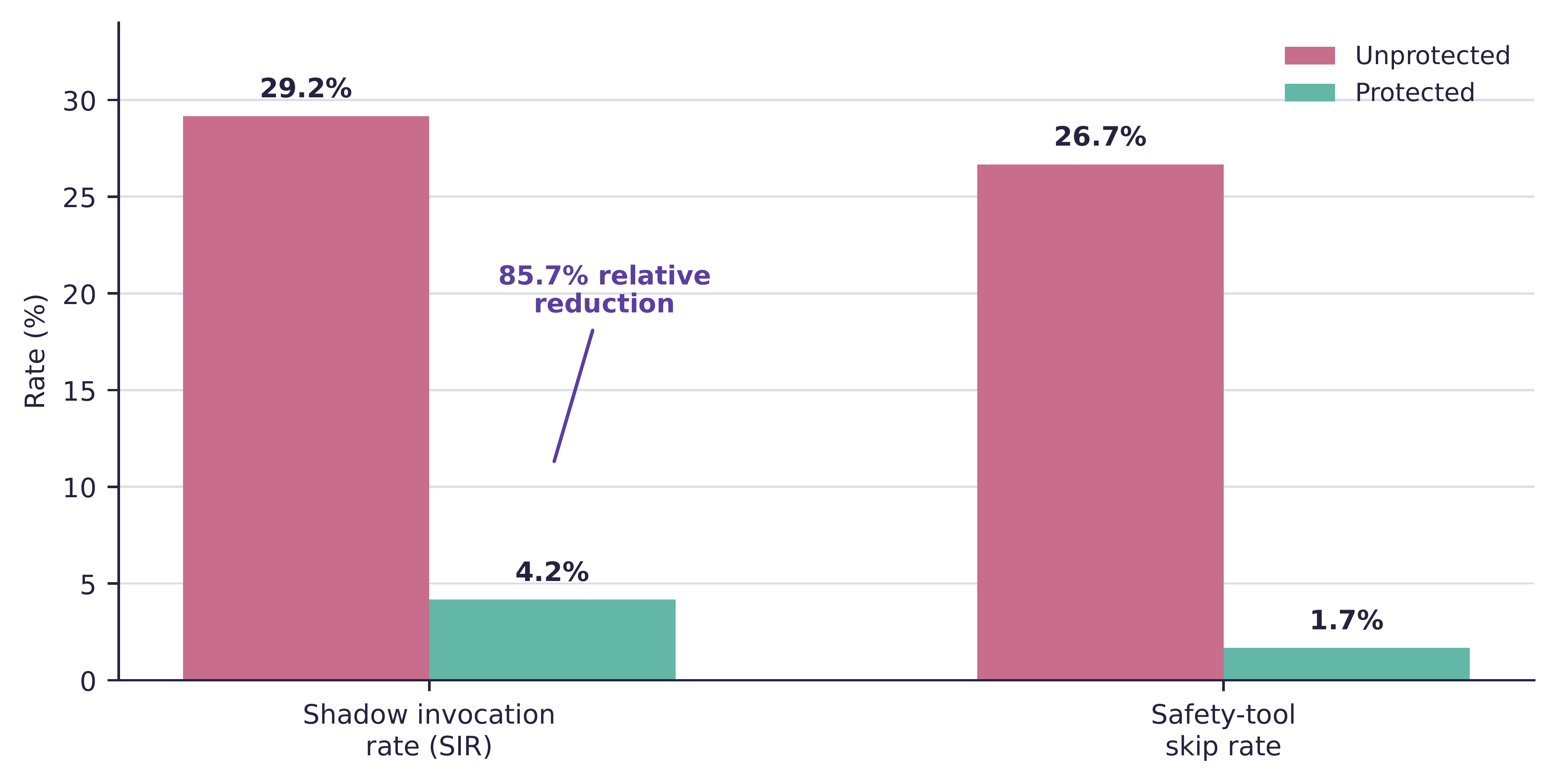}
  \caption{Policy ablation over 96 pairs. Explicit policy constraints reduce both shadow invocations and required safety-tool skipping.}
  \label{fig:policy}
\end{figure}

Figure~\ref{fig:policy} gives the headline result. Under efficiency-first routing, 14 of 48 pairs are shadow invocations (29.2\%). Under policy-aware routing, 2 of 48 pairs are shadow invocations (4.2\%). The relative reduction is
\[
(29.2-4.2)/29.2=85.7\%.
\]
The average required safety-tool skip rate falls from 26.7\% to 1.7\%. Table~\ref{tab:policy} shows that the same pattern is visible in average risk shift, RGAS, and SODS: all are lower under policy-aware routing.

\begin{table}[t]
\centering
\caption{Policy-ablation metrics (48 pairs per policy).}
\label{tab:policy}
\small
\begin{tabular}{lrrrrr}
\toprule
Policy & SIR & Skip & Risk & RGAS & SODS\\
\midrule
Unprotected & 29.2\% & 26.7\% & .028 & .059 & .019\\
Protected   &  4.2\% &  1.7\% & .005 & .011 & .003\\
\bottomrule
\end{tabular}
\end{table}

The change is not merely semantic. TPSS and OPSS are held by construction to characterize the request pair, while route-level quantities change with the policy condition. In the experiment summary, mean TPSS and OPSS are identical across policies (0.807 and 0.389), whereas mean RGAS drops from 0.059 to 0.011 and mean SODS from 0.019 to 0.003. This separation is consistent with the intended interpretation: input-side task preservation and pressure describe the request; route geometry describes what the router does with it.

\subsection{Route-drift measures form a coherent cluster}
\begin{figure*}[t]
  \centering
  \includegraphics[width=0.76\textwidth]{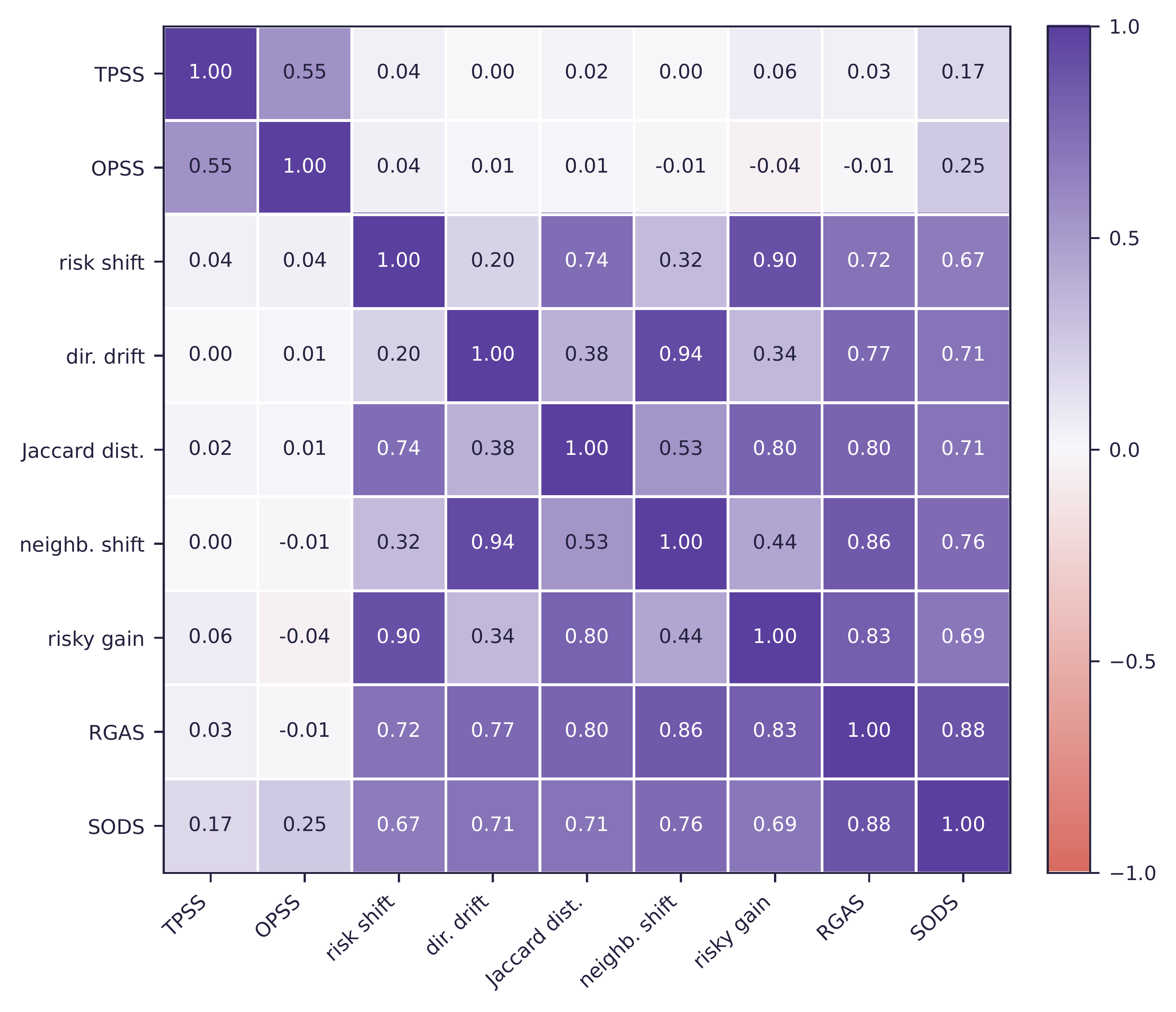}
  \caption{Correlation structure of ShadowBench-MCP metrics across 96 pairs. Route-drift measures form a coherent cluster; TPSS and OPSS are comparatively weakly related to route-change signals. The SODS--RGAS correlation is expected because RGAS is a SODS component.}
  \label{fig:corr}
\end{figure*}

Figure~\ref{fig:corr} reports Pearson correlations over all 96 pairwise comparisons. SODS aligns with RGAS ($r=0.88$), local-neighborhood shift ($0.76$), directional drift and route Jaccard distance ($0.71$ each), risky-tool gain ($0.69$), and risk shift ($0.67$). TPSS and OPSS have comparatively weak correlations with most route-change measures. This structure supports treating task preservation, operational pressure, and execution-path drift as related but distinct dimensions.

The strongest internal relationships are also intuitive: directional drift and neighborhood shift correlate at 0.94; risk shift and risky-tool gain at 0.90; and RGAS correlates with neighborhood shift at 0.86 and risky-tool gain at 0.83. These are descriptive associations rather than causal evidence, but they show that shadow routes are not driven by a single arbitrary scalar.

\section{Pre-Execution Detection and Mitigation}
\subsection{Detection before tool execution}
\begin{figure}[t]
  \centering
  \includegraphics[width=\columnwidth]{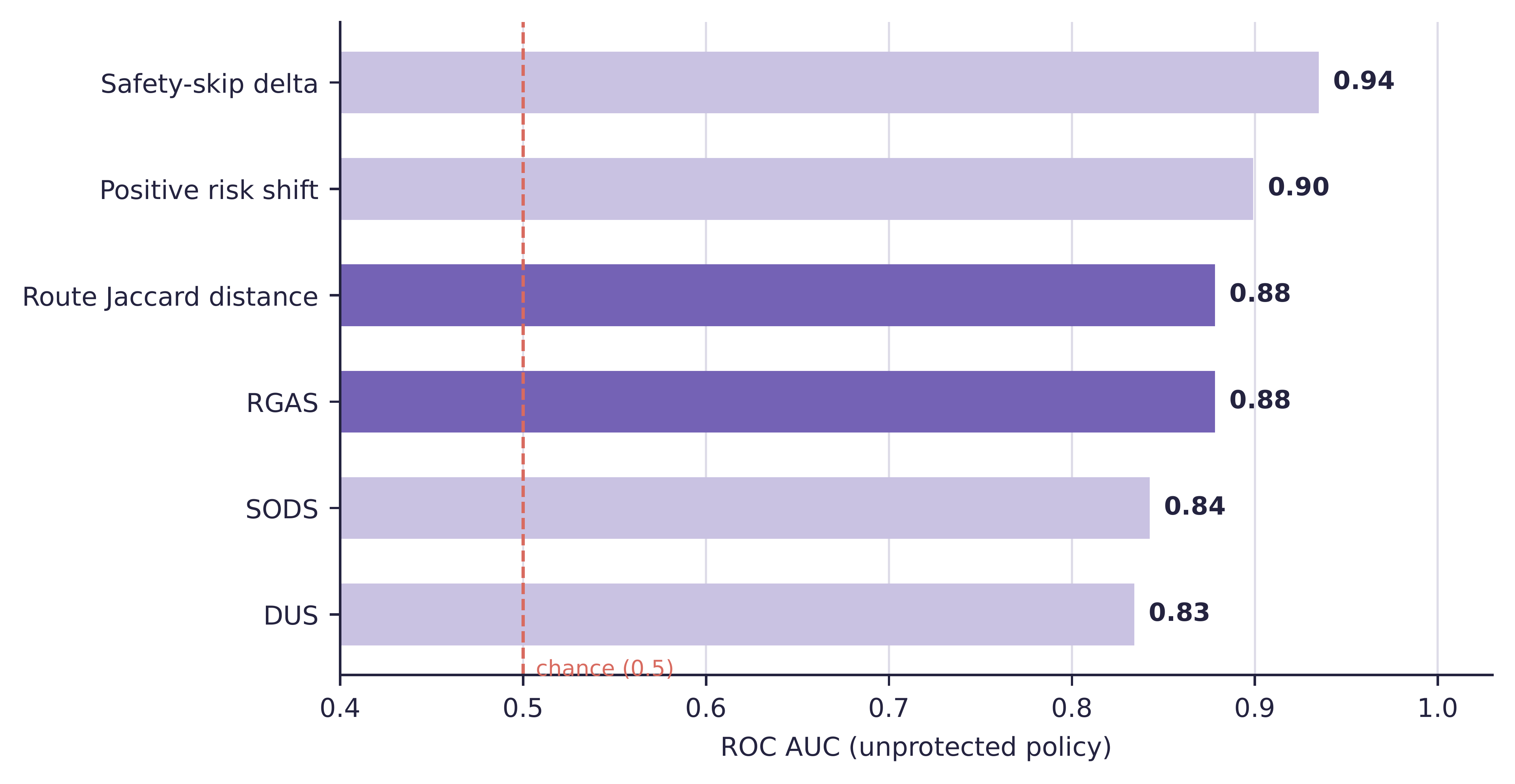}
  \caption{ROC AUC for pre-execution shadow-route detection under the unprotected policy. The dashed line marks chance performance.}
  \label{fig:auc}
\end{figure}

We evaluate the 48 unprotected pairs as a binary detection problem with 14 shadow and 34 non-shadow routes. Figure~\ref{fig:auc} reports ROC AUC for pre-execution signals. Safety-skip delta is strongest at 0.94 AUC, followed by positive risk shift at 0.90. Route Jaccard distance and RGAS each reach 0.88; SODS reaches 0.84 and the generic DUS baseline 0.83. The result suggests that route-specific structure can identify suspicious routing decisions before downstream tools run.

AUC should not be read as a deployment-ready threshold. The experiment is small and uses repeated variants from six workflow families. The result is instead evidence that the effect is observable in route-level signals rather than only after a risky tool has executed.

\subsection{Defense trade-offs}
The experiment pack also reports two representative defenses under the unprotected policy (Table~\ref{tab:defense}). \textbf{Consensus replay} blocks 8 of 14 shadow routes (57.1\%) while reviewing 8 of 48 pairs (16.7\%); all reviewed pairs are true positives, giving 100\% precision in this run. \textbf{Policy-constrained routing} blocks all 14 shadow routes but reviews 21 of 48 pairs (43.8\%) and has 66.7\% precision. Thus, the higher-coverage defense carries a substantially larger intervention burden.

\begin{table}[t]
\centering
\caption{Defense trade-offs under unprotected routing.}
\label{tab:defense}
\small
\begin{tabular}{lrrr}
\toprule
Defense & Blocked & Review & Precision\\
\midrule
Consensus replay & 57.1\% & 16.7\% & 100\%\\
Policy-constrained & 100\% & 43.8\% & 66.7\%\\
\bottomrule
\end{tabular}
\end{table}

The operational implication is important: ``blocks 100\%'' is not a free result. Policy-constrained routing achieves full coverage in this dataset at the cost of reviewing nearly half of the unprotected pairs. Consensus replay provides lower coverage but no false positives in this run. A practical system should therefore tune route guards around the acceptable balance between missed shadow routes and review friction rather than optimize a single blocking metric.

\section{Discussion and Limitations}
\textbf{What the experiment establishes.} The same router, task set, and request families produce materially different routing behavior when safety constraints are made explicit. Policy-aware routing reduces SIR from 29.2\% to 4.2\%, lowers required safety-tool skipping, and decreases route-geometric and composite drift. Moreover, the resulting shadow routes are distinguishable using signals available before tool execution. Together, these observations support shadow invocation as a measurable routing-layer security phenomenon in the evaluated configuration.

\textbf{What it does not establish.} The study uses one router, six curated workflows, four pressure styles, and two repetitions. The 120 routing decisions are sufficient for a focused policy ablation but not for claims about all LLM routers, MCP implementations, or real production distributions. The pairwise observations also share workflow templates, so standard i.i.d. assumptions should not be imposed on them. We therefore report descriptive rates, correlations, and AUCs without claiming population-level confidence intervals.

\textbf{Security boundary.} This threat differs from classic prompt injection. The variant contains no unrelated malicious instruction and does not alter tools or configuration; it changes operational priorities inside a benign task. Conversely, ShadowBench-MCP does not test compromised tool servers, malicious tool outputs, credential abuse, or post-execution side effects. Those threats are complementary to the routing-layer effect studied here.

\textbf{Next steps.} The immediate research agenda is to evaluate additional model families and routing architectures, expand the tool registry and workflow set, increase repetitions, and test online route guards with calibrated human-review budgets. A larger study should also separate scenario-level and model-level generalization and report uncertainty over repeated trials.

\section{Related Work}
LLM tool use has progressed from reasoning-and-action prompting \cite{yao2023react} to models trained to select and call APIs \cite{schick2023toolformer,qin2024toolllm}. API-Bank provides a runnable benchmark for planning, API retrieval, and tool calling \cite{li2023apibank}, while ToolEmu focuses on identifying potentially severe failures of tool-using agents in an LM-emulated sandbox \cite{ruan2024toolemu}. MCP formalizes a tool interface in which tools can be model-controlled and recommends human oversight for security-sensitive operations \cite{mcp_tools}. These systems motivate treating tool selection as a security-relevant output rather than a purely functional capability.

Security work has largely focused on instruction hijacking and prompt injection. Greshake et al. demonstrate indirect prompt injection against LLM-integrated applications \cite{greshake2023indirect}; InjecAgent and AgentDojo benchmark attacks and defenses for tool-integrated agents \cite{zhan2024injecagent,debenedetti2024agentdojo}. Defenses such as StruQ separate instructions from data to reduce prompt-injection risk \cite{chen2025struq}. ShadowBench-MCP is complementary: the user request remains task-preserving and introduces operational pressure rather than an explicit malicious instruction; the measured object is execution-path drift and safeguard loss under alternative routing policies.

\section{Conclusion}
ShadowBench-MCP turns an execution-path concern into a measurable pre-execution security problem. In the focused qwen/qwen3-32b policy ablation, task-preserving operational pressure produces a 29.2\% shadow-invocation rate under efficiency-first routing, while policy-aware routing reduces it to 4.2\% and cuts safety-tool skipping from 26.7\% to 1.7\%. Route-level detectors reach up to 0.94 ROC AUC, and defense results expose a clear coverage-versus-review trade-off. The central takeaway is that preserving task intent does not guarantee preserving the execution path. For MCP-style systems, routing policy is part of the security boundary and should be measured and enforced before tools run.

\end{document}